\documentclass{article}
\usepackage{spconf,amsmath,graphicx}

\usepackage{enumitem}
\setlist{nosep, leftmargin=14pt}

\usepackage{mwe} 
\usepackage{hyperref}
\usepackage{booktabs}
\usepackage{multirow}

\title{Transformer-based Model for Oral Epithelial Dysplasia Segmentation}
\name{
Adam J Shephard$^{1}$,
Hanya Mahmood$^{2}$,
Shan E Ahmed Raza$^{1}$,
Anna Luiza Damaceno Araujo$^{3}$, \\
\textit{Alan Roger Santos-Silva}$^{3}$,
\textit{Marcio Ajudarte Lopes}$^{3}$,
\textit{Pablo Agustin Vargas}$^{3}$,
\textit{Kris McCombe}$^{4}$, \\
\textit{Stephanie Craig}$^{4}$,
\textit{Jacqueline James}$^{4}$,
\textit{Jill Brooks}$^{5}$,
\textit{Paul Nankivell}$^{5}$,
\textit{Hisham Mehanna}$^{5}$, \\
\textit{Syed Ali Khurram}$^{2*}$,
\textit{Nasir M Rajpoot}$^{1*}$
}

\address{
$^{1}$ Tissue Image Analytics Centre, Department of Computer Science, University of Warwick, UK \\
$^{2}$ School of Clinical Dentistry, University of Sheffield, UK \\
$^{3}$ Oral Diagnosis Department, Piracicaba Dental School University of Campinas, S\~ao Paulo, Brazil \\
$^{4}$ Precision Medicine Centre, Queen’s University Belfast, UK \\
$^{5}$ Institute of Head and Neck Studies and Education, University of Birmingham, UK \\
$^{*}$ Joint co-senior authorship
}
\begin{document}
%
\maketitle
\begin{abstract}
Oral epithelial dysplasia (OED) is a premalignant histopathological diagnosis given to lesions of the oral cavity. OED grading is subject to large inter/intra-rater variability, resulting in the under/over-treatment of patients. We developed a new Transformer-based pipeline to improve detection and segmentation of OED in haematoxylin and eosin (H\&E) stained whole slide images (WSIs). Our model was trained on OED cases (n = 260) and controls (n = 105) collected using three different scanners, and validated on test data from three external centres in the United Kingdom and Brazil (n = 78). Our internal experiments yield a mean F1-score of 0.81 for OED segmentation, which reduced slightly to 0.71 on external testing, showing good generalisability, and gaining state-of-the-art results. This is the first externally validated study to use Transformers for segmentation in precancerous histology images. Our publicly available model shows great promise to be the first step of a fully-integrated pipeline, allowing earlier and more efficient OED diagnosis, ultimately benefiting patient outcomes.

\end{abstract}
\begin{keywords}
Oral Epithelial Dysplasia, Segmentation, Tranformer, Computational Pathology, Histopathology
\end{keywords}
\section{Introduction}

Oral epithelial dysplasia (OED) presents a significant challenge in the realm of head \& neck pathology, where accurate diagnosis and early detection are paramount for effective intervention and the prevention of malignant progression \cite{Speight2018}. OED is a premalignant histopathological diagnosis encompassing various lesions of the oral mucosa, typically manifesting as white (leukoplakia), red (erythroplakia) or mixed (red-white) lesions \cite{Speight2018}. Accurate diagnosis and early detection of OED are crucial for effective intervention and prevention of malignant progression. However, the current manual assessment of H\&E-stained sections of oral tissue slides, the gold standard in OED diagnosis, suffers from low throughput and susceptibility to intra-/inter-observer variability \cite{Speight2018, Kujan2007}.

To address these challenges and enhance the diagnosis and management of OED, there is a growing interest in leveraging advanced technologies, particularly deep learning, which has seen extensive use in medical image analysis over the last decade \cite{Litjens2017, Madabhushi2016}. Concurrently, Transformers have captured widespread attention in recent years due to their successful application in various domains, including natural language processing and computer vision tasks, such as classification \cite{he2023transformers}. A typical Transformer encoder comprises three fundamental components: a multi-head self-attention (MSA) layer, a multi-layer perceptron (MLP), and layer normalisation (LN). The inclusion of the MSA layer is particularly noteworthy as it empowers Transformers to capture long-range dependencies, rendering them a promising choice for semantic segmentation in the context of medical images \cite{Chen2021, cao2022}. 
While Transformers have demonstrated their potential to mitigate some of the constraints associated with convolutional neural networks (CNNs), their utilization in histological applications has been primarily limited to classification tasks, with semantic segmentation left relatively unexplored. This raises the question of whether Transformers can be harnessed for segmentation of histological images.

\begin{table*}[ht!]
    \centering
    \caption{Internal testing results with different loss functions and patch sizes/resolutions.}\label{tab:patch}
    \begin{tabular}{lccccccc}
        \hline
        \multirow{2}{*}{Loss} & \multirow{2}{*}{Patch Size} & \multirow{2}{*}{Res. (mpp)} & \multicolumn{3}{c}{OED cases} & Controls \\
        \cline{4-7}
        &&& F1 & Recall & Prec. & Spec. \\
        \hline
        Dice + CE & 256 & 1.0 & 0.794	& 0.824	& 0.767	& 0.998 \\
        Dice + CE & 512 & 0.5 & 0.781	& 0.792	& 0.771	& 0.999 \\
        \textbf{Dice + CE} & \textbf{512} & \textbf{1.0} & \textbf{0.807}	& 0.844	& 0.773	& 0.997 \\
        \hline
        Dice & 512 & 1.0 & 0.795 & \textbf{0.852} & 0.746 & 0.996 \\
        Jaccard & 512 & 1.0 & 0.000 & 0.000 & 0.000 & \textbf{1.000} \\
        CE & 512 & 1.0 & 0.805 & 0.834 & \textbf{0.778} & 0.998 \\
        Jaccard + CE & 512 & 1.0 & 0.784 & 0.828 & 0.744 & 0.996 \\
        \hline
    \end{tabular}
\end{table*}

In this study, we apply a Transformer-based model to a
comprehensive OED dataset for dysplasia segmentation, set-
ting a new standard in the field.  Our model is built on the Trans-UNet architecture \cite{Chen2021}, and is specifically designed for segmenting dysplastic regions in H\&E-stained whole slide images (WSIs) of oral tissue. We believe that the application of cutting-edge, state-of-the-art (SOTA) deep learning techniques, such as Transformer-based architectures, holds the potential to significantly improve the accuracy and efficiency of OED diagnosis. We rigorously evaluate the performance of our model by comparing it to other SOTA methods, and demonstrate its robustness and generalisability by extending our evaluation to include cases from three external and international centres: Birmingham (UK), Belfast (UK) and S\~ao Paulo (Brazil). We have open-sourced our model inference pipeline to facilitate broader research and application (\url{https://github.com/adamshephard/oed_inference}).

\section{Method}

\subsection{Study Data}

\subsubsection{Training Data}
The training dataset comprised a retrospective sample of histology tissue sections collected (dating 2008 to 2016) from the Oral and Maxillofacial Pathology archive at the School of Clinical Dentistry, University of Sheffield, UK. New tissue sections of the selected cases were cut (4 $\mu$m thickness) from formalin fixed paraffin embedded (FFPE) blocks and stained with H\&E. The dataset comprised 260 slides with a histological diagnosis of OED, and 105 non-dysplastic (control) slides. Slides were scanned at 40$\times$ objective power with either a NanoZoomer S360 (Hamamatsu Photonics, Japan; 0.2258 mpp), an Aperio CS2 (Leica Biosystems, Germany; 0.2520 mpp), or a Pannoramic 1000 (3DHISTECH Ltd., Hungary; 0.2426 mpp) slide scanner to obtain digital WSIs. Exhaustive delineation of ROIs representing dysplastic epithelium in OED slides, and normal epithelium in controls slides, was performed using QuPath \cite{Bankhead2017}. 

\subsubsection{External Testing Data}

For the external testing of the models generated in this study, we recruited OED cases from three external centres: (i) Queen’s University Belfast, UK; (ii) Institute of Head and Neck Studies and Education, Birmingham, UK; and (iii) Piracicaba Dental School, Brazil. 30 OED cases were collected from Belfast, 30 from Birmingham and 18 from Brazil. The Birmingham and Belfast slides were scanned at 40$\times$ objective power using a Pannoramic 250 (P250, 3DHISTECH Ltd., Hungary; 0.1394 mpp), and an Aperio AT2 (Leica Biosystems, Germany; 0.2529 mpp) scanner, respectively. The Brazil cases were scanned at 20$\times$ objective power, by an Aperio CS (Leica Biosystems, Germany; 0.4928 mpp) scanner. Owing to the limited size of the these datasets we combined them into a single multi-institutional test set consisting of 78 OED cases. Exhaustive delineation of dysplastic ROIs in the epithelium in all cases was performed.

\subsection{Network Architecture and Implementation}

\begin{figure}[t]
\centering
    \includegraphics[width=1.0\columnwidth]{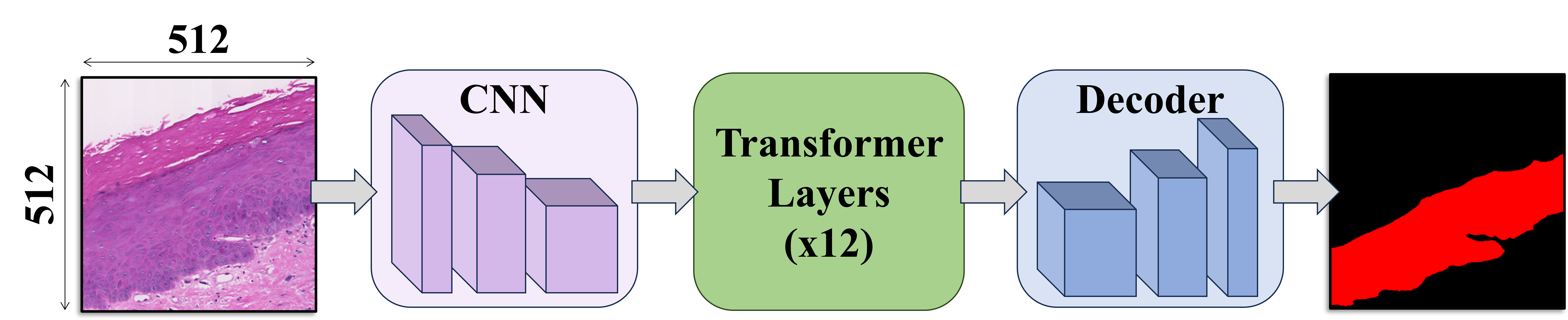}
    \caption{Network architecture of the TransUNet model.}\label{fig:network}
\end{figure}

We present a new model for OED segmentation, based on the TransUNet \cite{Chen2021} architecture (see Fig. \ref{fig:network}). This is a hybrid model, that uses a CNN (ResNet50 \cite{he2016}) as a feature extractor. $1\times1$ patches are then extracted from the feature maps and used for patch embedding for the Transformer layers. Finally, a cascaded upsampler is used as a decoder. This allows feature aggregation through skip-connections, thus leveraging the high-resolution CNN feature maps in the decoding path.

Our model takes an input RGB image of size $512\times512$ (at 1.0 micron per pixel, mpp, resolution) and outputs a dysplasia segmentation map.
For post-processing, we performed morphological closing/opening, and removal of small objects and holes.
We first tested the proposed model over varying patch sizes, resolutions, and loss functions. To aid our model in generalising to unseen domains, we tested its performance based on various domain generalisation (DG) techniques. Methods we employed included: weighted sampling (WS), stain augmentation (SA), and domain adversarial training (DA) \cite{ganin2016}. Following this, we compared our model against other state-of-the-art deep learning models for semantic segmentation, including Swin-UNet \cite{cao2022}, U-Net \cite{Ronneberger2015} (ResNet-50 \cite{he2016} backbone), Efficient-UNet \cite{Baheti2020} (Efficient-Net-B7 backbone), DeepLabV3+ \cite{Chen2018} (ResNet-101 backbone), and HoVer-Net+ (\cite{Shephard2021}; segmentation decoder alone). All of these models were trained based on their default parameters, and pretrained on ImageNet. 

We trained all models in two phases. We trained the decoders for 20 epochs first, before training the entire network for 30 epochs second. The Adam optimizer was used with a learning rate that decayed initially from 10\textsuperscript{-4} to 10\textsuperscript{-5} after 10 epochs, in both phases. We applied the following random data augmentations: flip, rotation, Gaussian blur, median blur, and colour perturbation. We additionally tested the effect of stain augmentation using the TIAToolbox \cite{Pocock2022} implementation of the Macenko method \cite{Macenko2009}. This has been shown previously to help counter scanner-induced domain-shift \cite{Aubreville2023, Jahanifar2021Mitosis, Jahanifar2022}. 

For internal testing, we split the dataset with a 80/20 split controlled for both scanner and OED grade. This resulted in 206 OED and 75 control slides in the training set, and 54 OED and 21 control slides in the testing set. A higher proportion of controls were kept in the test set to ensure a high specificity of OED segmentation in controls. An equal number of cases and controls were used from each scanner in the test set. We tessellated our WSIs and masks into smaller patches of size $512\times512$ (overlap of 184) pixels at 10$\times$ magnification (1.0 mpp), resulting in a total of 11,756 normal patches and 19,063 OED patches for model training/validation on the discovery cohort. For model testing, we report our evaluation metrics at the ROI level. 
Typically, each case/control had only one complete tissue section annotated. However, since some of the WSIs contained multiple tissue sections with annotations, this amounted to 66 OED ROIs and 23 control ROIs for testing. Each ROI encapsulated a whole tissue section. 

For external validation, we trained our models based on the Sheffield data, and tested on the 78 external OED cases. This resulted in a total of 6,341 OED patches for model validation. Since some of these WSIs contained multiple tissue sections with annotations, this totalled 87 OED ROIs. The external data only comprises OED cases (and no controls).

\subsection{Evaluation Metrics}
For OED cases, we report an F1-score, recall and precision, aggregated over all ROIs. For controls, we provide the model specificity, since a single false positive pixel, would result in F1, recall, and precision values of 0; thus not giving an accurate representation of the model performance.

\section{Experiments and Results}

\begin{table}[t]
    \centering
    \caption{Internal testing on the OED cases and controls, whilst testing domain generalisation techniques. }\label{tab:dom-gen}
    \begin{tabular}{lccccc}
    \hline
    \multirow{2}{*}{DG Method} & \multicolumn{3}{c}{OED cases} & Controls \\
    \cline{2-5}
    & F1 & Recall & Prec. & Spec. \\
    \hline
    WS & 0.798 & 0.839 & 0.760 & 0.998 \\
    SA \cite{Macenko2009} & 0.805 & \textbf{0.858} & 0.758 & 0.997 \\
    DA \cite{ganin2016} & 0.682 & 0.723 & 0.644 & 0.984 \\
    WS, SA & 0.802 & 0.851 & 0.758 & 0.997 \\
    WS, DA & 0.700 & 0.749 & 0.657 & 0.991 \\
    SA, DA & 0.735 & 0.774 & 0.701 & 0.992 \\
    WS, SA, DA & 0.699 & 0.725 & 0.655 & 0.988 \\
    \hline
    Proposed & \textbf{0.807} & 0.845 & \textbf{0.773} & \textbf{0.998} \\
    \hline
    \end{tabular}
\end{table}

\begin{table}[t]
    \centering
    \caption{Comparative experiments for internal testing.}\label{tab:in-comp}
    \begin{tabular}{lccccc}
    \hline
    \multirow{2}{*}{Model} & \multicolumn{3}{c}{OED cases} & Controls \\
    \cline{2-5}
    & F1 & Recall & Prec. & Spec. \\
    \hline
    U-Net \cite{Ronneberger2015} & 0.775 & 0.796 & 0.755 & 0.996 \\
    HoVer-Net+ \cite{Shephard2021} & 0.789 & 0.827 & 0.754 & 0.996 \\
    DeepLabV3+ \cite{Chen2018} & 0.802 & 0.817 & \textbf{0.788} & 0.998 \\
    Efficient-UNet \cite{Baheti2020} & 0.790 & 0.834 & 0.751 & 0.998 \\
    Swin-UNet \cite{cao2022} & 0.795 & 0.845 & 0.750 & 0.997 \\
    \hline
    Proposed & \textbf{0.807} & \textbf{0.845} & 0.773 & \textbf{0.998} \\
    \hline
    \end{tabular}
\end{table}

\begin{table}[!t]
    \centering
    \caption{Comparative experiments for external testing.}\label{tab:external-testing}
    \begin{tabular}{lccc}
    \hline
    Model & F1 & Recall & Prec. \\
    \hline
    U-Net \cite{Ronneberger2015} & 0.685 & 0.694 & 0.676 \\
    HoVer-Net+ \cite{Shephard2021} & 0.668 & 0.719 & 0.623 \\
    DeepLabV3+ \cite{Chen2018} & 0.704 & 0.704 & \textbf{0.705} \\
    Efficient-UNet \cite{Baheti2020} & 0.700 & \textbf{0.777} & 0.638 \\
    Swin-UNet \cite{cao2022} & 0.680 & 0.728 & 0.638 \\
    \hline
    Proposed & \textbf{0.708} & 0.764 & 0.660 \\
    Proposed (SA) & \textbf{0.708} & 0.744 & 0.676 \\
    \hline
    \end{tabular}
\end{table}

\begin{figure*}[t]
\centering
    \includegraphics[width=1.0\textwidth]{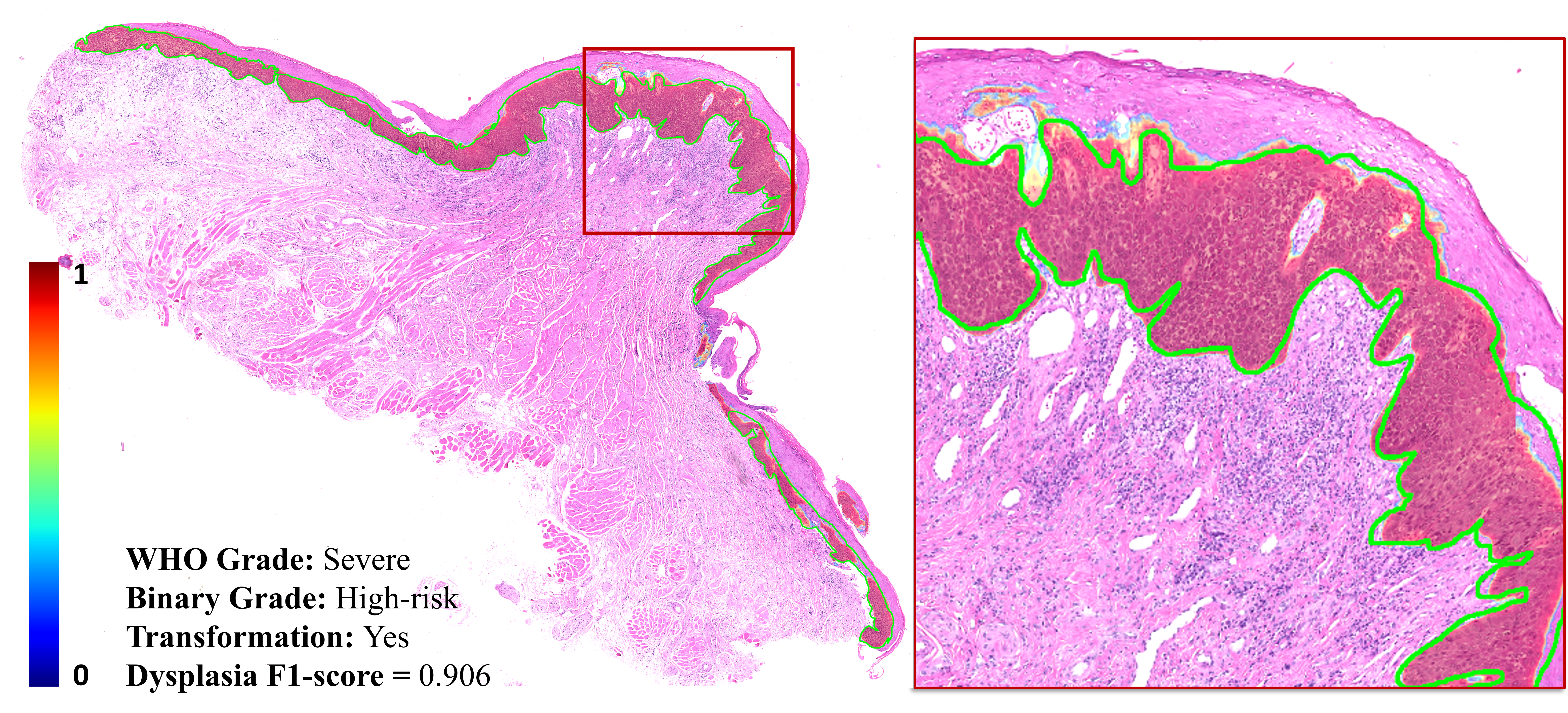}
    \caption{Dysplasia segmentation heatmap for a case graded as Severe (WHO Grade) and High-risk (Binary Grade), that transformed to cancer. The green line is the ground truth dysplasia segmentation by the pathologist. The red box shows a zoom in of one the most dysplastic areas.}\label{fig:seg}
\end{figure*}
We first tested the performance of our model over differing patch sizes (and resolutions), and loss functions; where we found a patch size of $512\times512$ at 1.0 mpp, with a combined Dice and cross-entropy loss function to be best (see Table \ref{tab:patch}). 
Next, we tested the proposed model when comparing the incorporation of various domain generalisation techniques (see Table \ref{tab:dom-gen}). These techniques yielded no improvement in performance on internal testing, with domain adversarial training hindering performance. Stain augmentation improved specificity on controls, with a slight reduction in F1-score. We suggest that these techniques were not beneficial on internal testing as slides from all three scanners were present in both the training and testing set. Instead, techniques such as stain augmentation may be more beneficial for external testing.

We compared our model to other state-of-the-art methods in Table \ref{tab:in-comp}. Here, we see the superiority of the proposed model (F1 = 0.81) when compared to all other models. DeepLabV3+ was the closest performing model (F1 = 0.80), with U-Net being worst (F1 = 0.78). We additionally provide a dysplasia heatmap for a severe OED case (see Fig. \ref{fig:seg}), showing our model's accurate segmentation.
The proposed model generalised well on external testing, gaining an F1-score of 0.71, and a high recall (see Table \ref{tab:external-testing}). Stain augmentation did not appear to improve the model F1-score; however, it did make the model more precise. We provide the comparative model results in Table \ref{tab:external-testing}, showing our proposed model to be best.

\section{Discussion and Conclusion}

In this study, we presented a Transformer-based model tailored for OED segmentation, and introduced the most extensive and diverse OED dataset to date. This dataset comprises 338 OED slides from four global centres, scanned using six different digital slide scanners, along with 105 control slides. 
Our study represents the first successful application of Transformer-based architectures for semantic segmentation in head \& neck histology images. Our model's architecture, featuring the Transformer's self-attention mechanism, enables it to capture long-range dependencies, making it well-suited for segmentation in complex medical images. Our model achieved remarkable performance, consistently outperforming other SOTA deep learning models. This underlines the technical prowess of Transformer-based architectures in tackling challenging medical image segmentation tasks.



We found one other study to perform OED segmentation \cite{liu2022}. This study focussed on moderate/severe OED cases, where dysplasia is more pronounced, and achieved an F1-score of 0.64 for segmentation at the patch-level. In comparison, all of our metrics are provided at the ROI-level, a harder task due to containing a higher variation of tissue type. Even so, we have clearly surpassed this performance on both internal (F1 = 0.81) and external testing (F1 = 0.71). 

A key technical achievement of our model is its robustness across diverse data sources. By training the model on slides from various scanners, we addressed a fundamental challenge in medical image analysis: domain shift \cite{jahanifar2023domain}. This ensured our model maintained its performance, even in the presence of variations introduced by different scanners/sites. Its ability to generalise well across external datasets, is a crucial indicator of its robustness and applicability in diverse clinical settings. 

As we move forward, the integration of the proposed model into clinical practice holds promise for enhancing the efficiency and reliability of OED diagnosis. 
Future research should focus on the seamless integration of this model into the assessment of individual slides for OED diagnosis and grading, enabling swifter and more objective treatment. Finally, the external validation of our models across multiple centres/scanners is a notable strength of this study. Future research could explore the application of the proposed model in even more diverse clinical settings and expand its utility to other histopathological tasks beyond OED. We suggest testing the method on other precancerous squamous lesions, such as laryngeal and cervical dysplasia, or even other types of dysplasia such as ductal carcinoma \textit{in situ}.


In conclusion, our research represents a substantial advancement in head \& neck pathology by providing a powerful publicly available model for OED segmentation, powered by a Transformer-based architecture. This technology demonstrates the transformative potential of computational pathology in improving the diagnosis and management of OED. As we address challenges and refine the model, deep learning is poised to play a vital role in enhancing the diagnosis of head \& neck precancerous lesions in the future. Finally, this works serves as a benchmark for future research into the use of Transformers for segmentation in histopathology images.
\pagebreak

\section{Compliance with Ethical Standards}

This study was performed in line with the principles of the Declaration of Helsinki. Ethical approval was granted by the NHS Health Research Authority West Midlands (18/WM/0335)

\section{Acknowledgements}

This work was supported by a Cancer Research UK Early Detection Project Grant (C63489/A29674), and a National Institute for Health Research grant (NIHR300904).

\bibliographystyle{IEEEbib}
\bibliography{references}

\end{document}